\documentclass[sigconf,nonacm]{acmart}

\AtBeginDocument{%
  }

\setcopyright{none}              
\renewcommand\footnotetextcopyrightpermission[1]{} 
\usepackage{amsmath}
\usepackage{graphicx}
\usepackage{booktabs}
\usepackage{multirow}
\usepackage{adjustbox}
\usepackage{rotating}
\usepackage{tikz}
\usepackage{xspace}
\usepackage{float}
\usepackage[table,xcdraw]{xcolor}

\newif\ifDEBUG
\DEBUGfalse  

\newif\ifBLINDED
\BLINDEDfalse  

\newif\ifARXIV
\ARXIVtrue  

\author{
{\rm Andrew T. Rozema}\\
Purdue University
\and
{\rm James C. Davis}\\
Purdue University
}

\newcommand{\ie}{\textit{i.e.},\xspace}
\newcommand{\eg}{\textit{e.g.},\xspace}
\newcommand{\etal}{\textit{et al.}\xspace}
\newcommand{\etals}{\textit{et al.}'s\xspace}

\usepackage{todonotes}
\usepackage{soul}

\ifDEBUG
    \newcommand{\DR}[1]{\todo[color=green,inline]{DR00:#1}}
    \newcommand{\JD}[1]{\todo[color=yellow,inline]{JD:#1}}
    \newcommand{\WJ}[1]{\todo[color=pink,inline]{WJ:#1}}

    \newcommand{\GKT}[1]{\todo[color=cyan,inline]{GKT:#1}}
    \newcommand{\KC}[1]{\todo[color=red,inline]{KC:#1}}
    
    \newcommand{\TODO}[1]{\hl{#1}}
\else
    \newcommand{\DR}[1]{}
    \newcommand{\JD}[1]{}
    \newcommand{\WJ}[1]{}
    \newcommand{\GKT}[1]{}
    \newcommand{\KC}[1]{}    
    \newcommand{\TODO}[1]{}
\fi

\usepackage{booktabs} 
\usepackage{colortbl}
\usepackage{amsmath}
\usepackage{amsfonts}
\usepackage{algpseudocode}
\usepackage{xspace}
\usepackage{multirow} 
\usepackage{balance} 
\usepackage{fancyvrb} 
\usepackage{url}
\definecolor{table_gray}{gray}{0.9}

\usepackage{cleveref}
\crefformat{section}{\S#2#1#3}
\crefname{figure}{Figure}{Figures}
\crefname{table}{Table}{Tables}
\crefname{listing}{Listing}{Listings}
\crefname{theorem}{Theorem}{Theorems}
\crefname{thm}{Theorem}{Theorems}
\crefname{lemma}{Lemma}{Lemmata}
\crefname{equation}{Eqt.}{Eqts.}
\crefformat{Grammar}{Grammar #1}

\usepackage{enumitem}
\usepackage{etoolbox}

\crefname{section}{Appendix}{Appendices}
\Crefname{section}{Appendix}{Appendices}

\newcommand{\myparagraph}[1]{\paragraph{#1}}
\renewcommand{\myparagraph}[1]{\paragraph{#1}}
\renewcommand{\myparagraph}[1]{\vspace{0.25em} \noindent \hspace{0.1cm} \textit{\textbf{#1:}}}

\usepackage{pifont} 
\usepackage{wasysym}

\newcommand{\cmarkgreen}{\textcolor{green!60!black}{\ding{51}}}  
\newcommand{\xmarkred}{\textcolor{red}{\ding{56}}}              

\usepackage{array}
\newcolumntype{P}[1]{>{\centering\arraybackslash}p{#1}}

\date{}
 
\title[Anti-Phishing Training (Still) Does Not Work]{Anti-Phishing Training (Still) Does Not Work: A Reproduction of Phishing Training Inefficacy Grounded in the NIST Phish Scale}

\begin{document}

\begin{abstract}
Social engineering attacks delivered via email, commonly known as phishing, represent a persistent cybersecurity threat leading to significant organizational incidents and data breaches.
Although many organizations train employees on phishing, often mandated by compliance requirements, the real-world effectiveness of this training remains debated.
Past work has demonstrated the ineffectiveness of training, but reproduction across different organizations, training approaches, and with a standardized threat assessment will help the generalizability of this phenomenon.

To contribute to evidence-based cybersecurity policy, we conducted a large-scale reproduction study ($N$=12,511) at a US-based financial technology firm.
Our design refined prior work by comparing training modalities in operational environments, applying NIST's standardized phishing difficulty measurement, and introducing novel organizational-level temporal resilience metrics.

Echoing prior work, training interventions showed no significant main effects on click rates ($p$=0.450) nor reporting rates ($p$=0.417), with negligible effect sizes.
However, we found that the NIST Phish Scale predicted user behavior, with click rates increasing from 7.0\% (easy lures) to 15.0\% (hard lures).
Our organizational-level resilience result was mixed:
  36-55\% of campaigns achieved ``inoculation'' patterns where reports preceded clicks, but training did not significantly improve organizational-level temporal protection.
Our results confirm the ineffectiveness of current phishing training approaches and offer a refined study design for future work. 

\ifARXIV
\fi
\end{abstract}

\ifARXIV
\fi
    
    \maketitle\
    
    \section{Introduction}

Phishing is a primary attack vector~\cite{greitzerExperimentalInvestigationTechnical2021} and regularly harms organizations through operational incidents, data breaches, supply chain attacks\cite{valsorda-compromise-survey},and financial losses~\cite{alkhalilPhishingAttacksRecent2021}.
For, in 2022--2023 campaigns attributed to the \emph{Scattered Spider} cluster
compromised 9,931 accounts in over 130 organizations~\cite{groupib2022oktapus,cisa2023scatteredspider}.
This campaign cost MGM Resorts alone an estimated \$100 million~\cite{reuters2024mgm100m}.
To prevent this class of attacks, organizations employ technical controls (\eg spam filters, attachment analyzers) complemented by cybersecurity awareness training~\cite{mengesContrastingSynergizingCISOs2024, hoUnderstandingEfficacyPhishing2025}.

Despite organizational and regulatory emphasis on anti-phishing training, its effectiveness is questioned by mounting empirical evidence~\cite{hoUnderstandingEfficacyPhishing2025, lain2022phishing, lain2024content} and employee sentiment~\cite{reevesGetRedhotPoker2021}.
Most pertinently, Lain \etal found that phishing training did not improve security outcomes across 19,000+ employees over years~\cite{lain2022phishing,lain2024content}.
These findings align with theoretical work by Volkamer \etal, who show fundamental limitations in human processing of security indicators~\cite{volkamer2020making}.

These prior studies support the general contention that phishing training is ineffective.
In this work we refine our understanding by addressing two specific knowledge gaps.
First, existing studies lack standardized measures of phishing difficulty, so we do not know whether training effectiveness varies by attack sophistication.
Second, most studies examine single training modalities rather than comparing different approaches systematically, so we do not know if some training designs are more effective than others.


To address these limitations, we conducted a large-scale reproduction study ($N$=12,511) at a US-based financial technology firm.\footnote{Our reproduction effort differs from \textbf{replication}, which would repeat identical procedures. Instead, we focus on testing the \textbf{generalizability} of the phenomenon of training ineffectiveness, varying organizational context and training approach.}
Our study design has three distinctive elements:
  (1) \textit{mandatory pre-phishing training} using two distinct training modalities (lecture-based vs. interactive);
  (2) \textit{systematic difficulty assessment} using the NIST Phish Scale~\cite{stevesCategorizingHumanPhishing2020,barrientosScalingPhishAdvancing2021}; and
  (3) a \textit{novel organizational temporal resilience measurement} through our Organizational Inoculation Index. 
Our design evaluates training effectiveness across standardized difficulty levels, and brings new insights into organizational cybersecurity resilience patterns to complement individual ones.

We confirm the ineffectiveness of anti-phishing training across multiple dimensions, but adds nuance to prior results.
Consistent with prior work,
neither training modality produced statistically significant improvements in 
  phish click reduction ($p$=0.450)
  nor
  phish reporting behavior ($p$=0.417),
  with negligible effect sizes indicating minimal practical value for individuals. 
However, the NIST Phish Scale successfully predicted user behavior (F(2, 12086)=41.415, p\ <\ 0.001, $\eta^2$=0.007), providing the first large-scale validation of this standardized difficulty measure.\footnote{The effect size is (statistically) small due to low base rates of phishing engagement. Practically, click rates doubled from 7.0\% to 15.0\% as lure difficulty increased.}
Beyond that, our Organizational Inoculation Index revealed substantial organizational resilience that operates independently of individual training effectiveness.
Across all experimental conditions, 36\%--55\% of template-group combinations demonstrated ``inoculation'' patterns where threat reports from one employee preceded threat interaction by others. 
\textbf{In summary, we contribute}:

\begin{enumerate}[leftmargin=1.5em, itemsep=0pt, topsep=0pt]
\item \textit{Reproduction of Training Ineffectiveness.} 
We reproduce the core finding of phishing training ineffectiveness. 

\item \textit{First Large-Scale Validation of NIST Phish Scale.}
We show the NIST Phish Scale 
predicts phishing lure difficulty. 

\item \textit{Introduction of the Organizational Inoculation Index.}
We introduce a novel metric for organizational temporal security.
We find substantial organizational resilience (36-55\% inoculation rates), independent of individual training effectiveness. 
\end{enumerate}

\vspace{0.1cm}
\noindent
\textbf{Significance:}
Cybersecurity practice and policy must be evidence-based.
For \textit{regulators}, our work adds more evidence that phishing training is ineffective. 
For \textit{organizations}, our results suggest realistic (low) expectations about training outcomes and highlight the importance of technical controls to assist humans. 
We also found that collective security behaviors may create measurable protective effects, emphasizing the value of user/IT feedback loops for real-time threat response.
For \textit{researchers}, we validate NIST's standardized difficulty measure and confirm the reproducibility of training ineffectiveness findings across organizational contexts.

\section{Background and Related Work}
\renewcommand{\arraystretch}{1.2} 
\rowcolors{2}{gray!10}{white}     

\begin{table*}[ht]
    \centering
    \footnotesize
    \small
    \caption{
        Summary of Prior Experiments on Phishing and Training Interventions.
        Studies are organized by research context (Lab vs. Real-World).
        \textit{Context}:
          ``Lab'' denotes research simulations, where subjects are not in their work roles.
          ``Real-World'' denotes studies where participants were employees acting in their job capacity.
        \textit{Complexity measure:}
          Whether the study controlled for the difficulty of the phishing tasks.
          \cmarkgreen\ indicates the use of an open control \JD{cite this term `open control' please} such as NIST Phish Scale,
          and
          \xmarkred\ indicates no control. 
        \textit{Hypotheses:} How these works informed the hypotheses in the present study (cf.~\cref{tab:HypothesisTable}).
          }
    \label{tab:phishing-study-review}
    \begin{tabular}{P{1.75cm}|cP{3.0cm}P{2.0cm}|P{2.51cm}P{2.70cm}|P{1.5cm}}
        \toprule
        \textbf{Work} & \textbf{Context} & \textbf{Population} & \textbf{Sample size} & \textbf{Complexity Measure} & \textbf{Compared Training Approaches?} & \textbf{Hypotheses} \\
        \midrule

                \cite{schoniYouKnowWhat2024} (2024) & Lab & University students & 96 & \xmarkred & \xmarkred & H3 \\

                \cite{canhamNotAllVictims2024} (2024) & Lab & University students & 117 & \cmarkgreen\ (Phish Scale) & \xmarkred & H4 \\



                \midrule

                \cite{caputoGoingSpearPhishing2014} (2014) & Real-world & Employees & 1,359 & \xmarkred & \cmarkgreen & H2, H4 \\

                \cite{carellaImpactSecurityAwareness2017} (2017) & Real-world & University students & 150 & \xmarkred & \cmarkgreen & H2, H3 \\

                \cite{sumnerExaminingFactorsImpacting2022} (2022) & Real-world & Employees & 119 & \xmarkred & \xmarkred & H2 \\
                
                \cite{lain2022phishing} (2022) & Real-world & Employees & 19,000+ & \xmarkred & \xmarkred & H2 \\
                \cite{hillmanEvaluatingOrganizationalPhishing2023} (2023) & Real-world & Finance employees & 5,000 & \xmarkred & \xmarkred & H2 \\
                
                \cite{lain2024content} (2024) & Real-world & Employees & 4,554 & \xmarkred & \cmarkgreen & H2, H3 \\
                
                \cite{doingAnalysisPhishingReporting2024} (2024) & Real-world & Banking employees & ``Thousands'' & \xmarkred & \xmarkred & H2 \\

                \cite{hoUnderstandingEfficacyPhishing2025} (2025) & Real-world & Healthcare employees & 19,500+ & \xmarkred & \xmarkred & H2 \\

        \midrule

        \rowcolor{yellow!20}
        \textbf{Ours} & Real-world & Fintech org. & 12,511 & \cmarkgreen\ (Phish Scale) & \cmarkgreen & --- \\ 

        \bottomrule
    \end{tabular}
\end{table*}

We review phishing attacks and defenses (\S\ref{subsec:phishing-defenses}),
  and
  prior empirical studies on phishing training (\S\ref{subsec:empirical-studies}).

\subsection{Phishing Attacks and Defenses}
\label{subsec:phishing-defenses}

Phishing attacks pose substantial risks to organizations, who erect both technical and human defenses against them.

\subsubsection{Phishing Attacks}
\label{subsec:PhishingAttackers}


In a phishing attack, attackers send messages (\textit{phishing lures}) to targets whom they wish to persuade to act against their own interests, and the interests of their organizations.
Attackers seek results such as sensitive information, account credentials, or the ability to control the victim's machine~\cite{alkhalilPhishingAttacksRecent2021, ferreiraPersuasionHowPhishing2019}. These attacks use compromised infrastructure, browser vulnerabilities\cite{10.1145/3589334.3645535}, look-alike domains, legitimate certificates \cite{kim2021sooel} and social engineering to evade technical defenses \cite{10.1145/3696410.3714678, 10.1145/3652963.3655042} and exploit human trust~\cite{alkhalilPhishingAttacksRecent2021, aleroudPhishingEnvironmentsTechniques2017}.
Traditionally, phishing attacks are divided into two classes~\cite{dasSoKComprehensiveReexamination2020}:
  mass phishing (form emails sent to many targets, often with a phishing kit \cite{10.1145/3696410.3714710})
and spear phishing (custom emails sent to specific targets).

\subsubsection{Automated Defenses}
\label{subsec:tech-defenses}

Many technical defenses combat attackers' behavior, such as
  SPF, DKIM, and DMARC to verify that mail comes unaltered from authorized infrastructure~\cite{rfc7208spf2014,rfc6376dkim2011,rfc7489dmarc2015};
  ARC for preserving authentication results across forwarding~\cite{rfc8617arc2019};
  BIMI for verified logos~\cite{bimi-group,bimi-ietf-guidance-2025};
  and
  secure email gateways (SEG) for content and header inspection and URL rewriting for time-of-click inspection~\cite{msft-safe-links-2025}.
After delivery, organizations employ continuous monitoring and feedback loops to adapt controls to evolving campaigns~\cite{mengesContrastingSynergizingCISOs2024}.
For an example of the dynamics~\cite{alkhalilPhishingAttacksRecent2021,aleroudPhishingEnvironmentsTechniques2017},
  security operations center (SOC) analysts triage user-reported messages, promote high-confidence indicators back into blocklists, and tune SEG policies based on observed false positives/negatives.


Appendix~\ref{sec:appendix:EmailControlsDetail} describes elements of the studied organization's email controls and feedback loops.

\subsubsection{Phishing Training as a Defense}
\label{subsec:human-focus}

Because phishing attacks compromise these technical defenses~\cite{butaviciusBreachingHumanFirewall2015, burnsSpearPhishingBarrel2019}, 
the human recipients play an important role in phishing defense~\cite{downsDecision2006, canfieldBetter2019,abroshanPhishingHappensTechnology2021, desoldaHuman2022}.
Training methods
range from lecture-style training involving passive videos and readings~\cite{marshallExploringEvidenceEmail2024} to simulation feedback providing immediate feedback after interaction with phishing simulations~\cite{carellaImpactSecurityAwareness2017, caputoGoingSpearPhishing2014}. 
Interactive training uses gamification or scenario-based learning to engage participants~\cite{bitrianGamificationWorkforceTraining2024, hartRiskioSeriousGame2020}, while just-in-time training alerts users in real time during risky behavior~\cite{franklinOptimisingNudgesBoosts2019}.

Anti-phishing trainings are often included in the cybersecurity awareness trainings ubiquitous in organizations.
Legal mandates drive this practice. 
For example:
  HIPAA requires security awareness training for healthcare organizations~\cite{hipaa-security-rule},
  and
  GDPR and PCI DSS encourage or mandate similar training~\cite{gdpr-article39,pci-dss}.
Industry standards reinforce these requirements~\cite{nist-csf}, \eg
  ISO 27001 calling for human security controls like awareness programs~\cite{iso-27001}.

\subsection{Empirical Studies of Phishing Training}
\label{subsec:empirical-studies} 

Our review and analysis of prior work is summarized in~\cref{tab:phishing-study-review}.
We contribute to the body of knowledge on phishing training by conducting another large-scale study situated in a real-world context (table columns 2-4), incorporating a complexity measure (column 5), and a comparison of multiple training approaches (column 6).
Our study tests hypotheses derived from prior work (final column).
The remainder of this section describes these contributions.
For general surveys of knowledge about human factors in phishing training, we refer the reader to~\cite{marshallExploringEvidenceEmail2024,franzSoK}.

\myparagraph{Lab vs. Real-world Studies}
Many early phishing training studies were conducted in controlled laboratory settings, where researchers benefit from a high degree of experimental control.
Studies of this sort have provided numerous insights, including that active learning and immediate feedback can improve short-term detection rates, and have explored how individual differences affect training outcomes~\cite{sumnerExaminingFactorsImpacting2022, marshallExploringEvidenceEmail2024}.
However, as noted by Franz \etal~\cite{franzSoK}, these controlled settings may lack ecological validity. In laboratory environments, participant samples are typically small and consist of homogeneous groups such as academics and students. 

In response, a growing number of studies have begun examining phishing training in real-world organizational environments. 
For example, Doing \etal~\cite{doingAnalysisPhishingReporting2024} analyzed phishing behaviors in bank employees, and Ho \etal~\cite{hoUnderstandingEfficacyPhishing2025} studied healthcare employees. 
Lain \etal has also conducted large-scale real-world measurements of training effectiveness~\cite{lain2022phishing,lain2024content}
Our study echoes some of these prior results in a different organizational context.

\myparagraph{Controlling for Lure Complexity}
These studies consistently report little effectiveness of anti-phishing training.
These works have not, however, incorporated a standardized method for assessing the difficulty of the phishing lures they use~\cite{hoUnderstandingEfficacyPhishing2025, doingAnalysisPhishingReporting2024}. 

The US NIST has proposed a scale for this purpose~\cite{stevesCategorizingHumanPhishing2020,dawkinsScalingNISTPhish2023}.
We summarize it here and give more details in~\Cref{sec:appendix-NISTPhishScale}.
The \textbf{NIST Phish Scale} categorizes lure difficulty along two dimensions.
\textit{Phishing Cues} represent observable errors or inconsistencies in the email that might alert users to its fraudulent nature, such as spelling mistakes, suspicious URLs, or formatting irregularities~\cite{canhamNotAllVictims2024}.
\textit{Premise Alignment} measures the relevance of the email to the recipient's organizational context~\cite{antoniouSystematicMethodExecute2015}.
Higher alignment messages with fewer cues are the most deceptive~\cite{butaviciusBreachingHumanFirewall2015}.
The premise of Phish Scale is to allow organizations to benchmark lure complexity and contextualize simulation outcomes~\cite{canhamPhishDerbyShoring2022}.
Building on early validation work~\cite{barrientosScalingPhishAdvancing2021,dawkinsScalingNISTPhish2023},
we contribute the first real-world measurement of the effect of lure complexity on subjects' interaction behaviors.

\myparagraph{Training Approaches}
Franz \etal~\cite{franzSoK} taxonomize five categories of phishing interventions:
  training, awareness, education, notifications, and design interventions.
Most studies have focused on training, which can be divided into lecture-style annual training and embedded training (\eg via ongoing phishing simulations from the organization's security team).
Like a few prior studies~\cite{caputoGoingSpearPhishing2014,carellaImpactSecurityAwareness2017,lain2024content}, we compare the effectiveness of two forms of phishing training. 

This study addresses those gaps by applying the NIST Phish Scale~\cite{dawkinsScalingNISTPhish2023} to evaluate training effectiveness across varied phishing difficulty levels in an enterprise deployment with over 12,000 participants.
In this study, our objectives are to:
(1) operationalize the NIST framework at scale,
(2) examine how difficulty influences detection and reporting across training modalities,
(3) provide comparative evidence on lecture-based versus interactive approaches,
and
(4) offer guidance for large-scale awareness efforts under conditions of modest effectiveness~\cite{hoUnderstandingEfficacyPhishing2025}. This enables evidence-based decision-making on phishing defense strategies that account for both training design and message complexity.

\section{Methodology}
\label{sec:Methods}

\begin{figure*}[ht]
    \centering
    \includegraphics[width=1\linewidth,trim=0 40 0 35,clip]{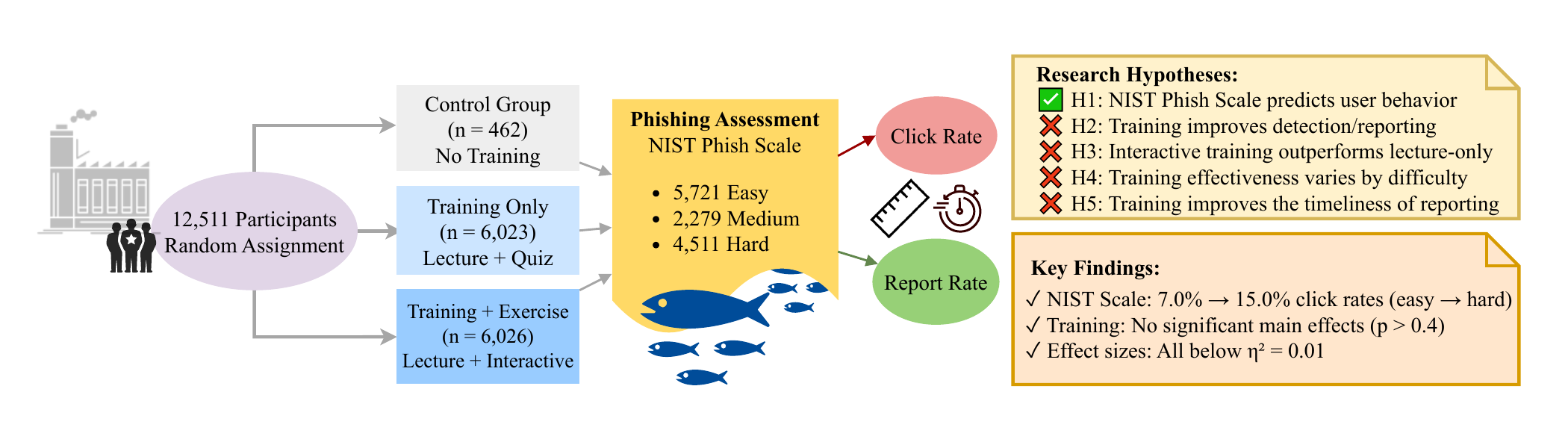}
    \Description{A diagram showing the study design. Participants are randomly assigned to different training modalities and receive phishing emails of varying difficulty; their click and report behaviors are measured and compared.}
    \caption{
    Illustration of our methodology.
    Our two-factor design randomly assigns different trainings to different subjects, and then randomly sends phishing lures of varying complexity to the subjects.
    We compared between-subject performance using a mix of standard and novel metrics to test our five hypotheses. 
    }
    \label{fig:MethodOverview}
\end{figure*}

To summarize the preceding analysis:
  phishing training is widely practiced, though recent evidence sheds doubt on its utility.
Few studies have rigorously assessed outcomes against standardized difficulty measures at sufficient scale for robust inference.
We aim to reproduce, testing hypotheses derived from prior works.

\subsection{Study Design}
\label{sec:Methods-Design}

Our study design is illustrated in~\cref{fig:MethodOverview}.
We employed a two-factor between-subjects experimental design~\cite{hillmanEvaluatingOrganizationalPhishing2023}.
Subjects were randomly assigned to one of three experimental conditions using our organization's training management system.
Each condition reflected the phishing training they received. 
Subjects were randomly selected, trained, and later received simulated phishing emails of varying complexity. 
We quantitatively compared subjects' performance in handling these phishing emails between the treatments. 

\myparagraph{Factor 1: Training}
The research team collaborated with a cybersecurity training vendor who produces comprehensive security awareness training materials, phishing simulation exercises designed to augment those materials, and implementation assistance for organizational deployment.
This partnership enabled access to professional training content and established phishing templates. 

The vendor offers two distinct training approaches for phishing.
The first approach, \textit{\ul{Treatment A}}, consists of lecture-based videos on the dangers of various social engineering attacks, including phishing recognition tactics, social engineering techniques, and organizational security protocols.
Comprehension is assessed through a quiz.
The second approach, \textit{\ul{Treatment B}}, consists of those videos followed by a series of interactive phishing exercises that present users with samples from a library of approximately 20 phishing templates.
These exercises guide users through identifying suspicious elements and security deficiencies in emails that roughly correspond to the cues assessed by the NIST Phish Scale, and reinforces appropriate response behaviors such as the ``Report Phish'' button in the email client~\cite{carellaImpactSecurityAwareness2017}.
Finally, we included a \textit{\ul{Control Group}} whose members received no cybersecurity training prior to the phishing simulation phase.
This group was included to establish baseline susceptibility measurements.\footnote{Control group participants received the same training materials after completion of the study to ensure organizational compliance with cybersecurity awareness requirements. The study timing dictated a small control group, as ACME Corp. required that most employees receive annual cybersecurity training for compliance purposes.}

\myparagraph{Factor 2: Phishing Lure Complexity}
After training, all subjects received one simulated phishing email, of varying complexity. 
To establish controlled and consistent evaluation of participant responses, we systematically assessed and categorized the difficulty of each simulated phishing email using a structured expert review process informed by the NIST Phish Scale framework~\cite{dawkinsScalingNISTPhish2023}. 

Phishing templates were selected from the vendor's library. 
We evaluated these templates using the official NIST Phish Scale User Guide methodology to ensure accurate difficulty classification.
For the \textit{Phishing Cues} dimension, raters assessed observable indicators that might alert users to fraudulent content, including spelling errors, suspicious URLs, formatting inconsistencies, and sender authenticity markers~\cite{canhamNotAllVictims2024}. 
For the \textit{Premise Alignment} dimension, raters evaluated how well each email's subject matter, sender, and content aligned with typical organizational communications that ACME Corp. employees receive in their regular workflow~\cite{abroshanPhishingHappensTechnology2021}.

The assessment team comprised three members of ACME Corp.'s IT security staff, each experienced in identifying real-world phishing attacks. 
Following Downs \etals advice, two team members independently evaluated each phishing email template~\cite{downsDecision2006}. 
Any disagreements regarding difficulty classification were discussed and resolved, incorporating a third rater as needed. 
This collaborative validation process ensured that our phishing simulation represented a controlled range of difficulty levels essential for evaluating training effectiveness across standardized complexity measures~\cite{barrientosScalingPhishAdvancing2021}.


Our final simulation campaign used nineteen phishing templates that we divided into Easy (high cues/low alignment --- 5721 emails), Medium (some cues, some alignment --- 2279 emails) and Hard (few cues/high alignment --- 4511 emails).
A full summary is given in~\cref{tab:phish-scale-assessment} in~\Cref{sec:appendix-Lures}.

\myparagraph{Simulation Deployment}
Phishing simulations were distributed among participants within a three-month window after training completion.
Each participant received one phishing email.
Templates were randomly assigned to ensure balanced distribution across training conditions.

\subsection{Subjects}
\label{sec:Methods-Recruitment}
\myparagraph{Study Context}
The subjects were full-time employees at a U.S.-based international financial technology (``fintech'') firm.
Subjects represented a diverse cross-section of the organization, including employees from finance, technology, operations, customer service, and administrative functions.


\myparagraph{Subject Recruitment}
This study leveraged organizational security training data collected as part of routine cybersecurity awareness training.
The training was introduced in official organizational communications emphasizing its importance for maintaining security compliance and protecting organizational assets.
This training was mandatory due to the organization's regulatory environment. 

Data were collected from Dec. 2024 -- Apr. 2025.
Participants had a one month window to complete their assigned training modules, with automated reminders sent at regular intervals.
Training completion was tracked and verified.

\myparagraph{Randomization}
Subjects were randomly assigned to each treatment.
To mitigate bias by organizational role or function, we used stratified sampling~\cite{sutterAvoidingHookInfluential2022} across departments to ensure representative distributions in each treatment group. 

\myparagraph{Completion Rates}
Our data came from the 12,511 participants who completed the full experimental protocol.
\cref{tab:recruitment-completion} shows the completion rates by treatment type.
Comparable completion rates across conditions (ranging from 87.3\% to 89.1\%) suggest our random sample was preserved between recruitment and completion. 


\rowcolors{2}{white}{gray!20}  
\begin{table}
\centering
\caption{
Recruitment pool and completion rates by condition.
}
\label{tab:recruitment-completion}
\footnotesize
\begin{tabular}{lcc}
\toprule
\textbf{Condition} & \textbf{Recruited} & \textbf{Completed} \\
\midrule
Treatment A (Lecture) & 6,758 & 6,023 (89.1\%) \\
Treatment B (Lecture + Exercises) & 6,764 & 6,026 (89.1\%) \\
Control Group & 529 & 462 (87.3\%) \\
\midrule
\textbf{Total} & \textbf{14,051} & \textbf{12,511 (89.0\%)} \\
\bottomrule
\end{tabular}
\end{table}



{
\renewcommand{\arraystretch}{1.50}
\rowcolors{2}{white}{gray!20}  

\begin{table*}[t]
\centering
\footnotesize
\small
\caption{
Summary of study hypotheses, their theoretical grounding, and relation to prior work.
The hypotheses address the effects of phishing email difficulty, training, and training modality on detection and reporting behaviors.
}
\label{tab:HypothesisTable}
\begin{tabular}{@{}p{0.30\textwidth} p{0.33\textwidth} p{0.30\textwidth}@{}}
\toprule
\textbf{Hypothesis} & \textbf{Theoretical and Empirical Basis} & \textbf{Relation to Prior Work} \\
\midrule
\textbf{H1--Phish Scale Effect:} \textit{Phishing emails with higher difficulty, as measured by the NIST Phish Scale, will have higher click-through rates than less difficult ones.} &
Grounded in the foundational NIST Phish Scale framework~\cite{dawkinsScalingNISTPhish2023} and validation studies~\cite{barrientosScalingPhishAdvancing2021}. Supported by phishing cue research~\cite{canhamNotAllVictims2024} and systematic email analysis~\cite{wrightPhishingSusceptibilityContext2023}. &
The NIST Phish Scale has been validated in lab settings~\cite{barrientosScalingPhishAdvancing2021}. We provide the first large-scale enterprise validation. \\
\textbf{H2--Training Impact (General):} \textit{Training will increase the detection and reporting of phishing emails.} &
Evidence of training efficacy, reduced click rates~\cite{carellaImpactSecurityAwareness2017}, improved detection accuracy~\cite{sumnerExaminingFactorsImpacting2022}, and short-term benefits~\cite{marshallExploringEvidenceEmail2024}.
Disputed~\cite{hoUnderstandingEfficacyPhishing2025}. &
Tests whether training effects from small-scale studies~\cite{carellaImpactSecurityAwareness2017, sumnerExaminingFactorsImpacting2022} replicate at scale, addressing Franz \etals call~\cite{franzSoK}. \\
\textbf{H3--Interactive Training Effect (Reporting):} \textit{Interactive components yield higher reporting rates.} &
Per Franz \etals analysis~\cite{franzSoK}, and evidence of effectiveness of active and feedback-rich training~\cite{carellaImpactSecurityAwareness2017, marshallExploringEvidenceEmail2024, schoniYouKnowWhat2024}. &
Lab findings show interactive training benefits~\cite{carellaImpactSecurityAwareness2017, marshallExploringEvidenceEmail2024}. We provide the first test at organizational scale. \\
\textbf{H4--Interaction Effect:} \textit{There will be an interaction between training modality and phishing difficulty, with interactive training more beneficial for easier lures.} &
Emerges from cognitive load and decision strategy research~\cite{downsDecision2006}, difficulty-aware detection patterns~\cite{wrightPhishingSusceptibilityContext2023}, and findings on human limitations~\cite{abroshanPhishingHappensTechnology2021}. &
Addresses the gap identified by Franz \etal~\cite{franzSoK} regarding interaction effects between training and phishing complexity, largely unexplored empirically. \\
\textbf{H5--Reporting Timeliness:} \textit{Trained users will report suspicious emails more quickly than untrained users.} &
Grounded in studies showing that training enhances threat recognition speed~\cite{carellaImpactSecurityAwareness2017}. 
Supported by evidence that security education improves decision-making efficiency~\cite{sumnerExaminingFactorsImpacting2022}. &
While individual reporting speed has been measured~\cite{lain2022phishing}, no studies have examined training effects on organizational-level reporting timeliness~\cite{stevesCategorizingHumanPhishing2020}. \\ 

\bottomrule
\end{tabular}
\end{table*}
}
\rowcolors{0}{}{} 

\subsection{Hypotheses}
\label{sec:Methods-Hypotheses}

Based on prior work on phishing training efficacy~\cite{carellaImpactSecurityAwareness2017, caputoGoingSpearPhishing2014, haneyComplianceImpactTracing2024, sumnerExaminingFactorsImpacting2022, franzSoK, hoUnderstandingEfficacyPhishing2025} and documented differences in risk perception~\cite{abroshanPhishingHappensTechnology2021, doingAnalysisPhishingReporting2024}, we formulated and tested five hypotheses.
\cref{tab:HypothesisTable} states each hypothesis, its basis, and its relation to prior work.
\textbf{\textit{We hypothesize that:}}

\begin{enumerate}[leftmargin=1.5em, itemsep=0pt, topsep=0pt]
\item \textbf{H1 (Phish Scale Effect):} \textit{Phishing emails with higher difficulty, as measured by the NIST Phish Scale, will have higher click-through rates than less difficult ones.}
\item \textbf{H2 (Training Impact--General):} \textit{Training will increase the detection and reporting of phishing emails.}
\item \textbf{H3 (Interactive Training Effect--Reporting):} \textit{Interactive components yield higher reporting rates than lecture-only training.}
\item \textbf{H4 (Interaction Effect):} \textit{There will be an interaction between training modality and phishing difficulty, with interactive training more beneficial for easier lures.}
\item \textbf{H5 (Reporting Timeliness):} \textit{Training will improve the timeliness of phishing reporting, with trained users reporting suspicious emails more quickly than untrained users.}
\end{enumerate}

\vspace{0.05cm}
\noindent
With respect to hypothesis novelty and prior testing:
  \textbf{H1} has been tested in small academic settings~\cite{barrientosScalingPhishAdvancing2021} but never at enterprise scale.
  \textbf{H2} and \textbf{H3} build on established training effectiveness research~\cite{carellaImpactSecurityAwareness2017, sumnerExaminingFactorsImpacting2022, marshallExploringEvidenceEmail2024} but applies these findings in a large-scale operational context using a standardized difficulty measure;
  \textbf{H4} is novel, addressing the research gap identified by Franz \etal~\cite{franzSoK} regarding interactions between training modality and phishing complexity, which has received minimal empirical attention in prior work.
  \textbf{H5} is also novel, directly addressing Steves \etal's research question about organizational temporal patterns~\cite{stevesCategorizingHumanPhishing2020}. 


\subsection{Metrics}
\label{sec:Methods-Metrics}

Our outcome measures were a mix of standard metrics for comparison to prior results~\cite{carellaImpactSecurityAwareness2017, caputoGoingSpearPhishing2014, lopezaguilarPhishingVulnerabilityPersonality2025} and novel metrics related to response timeliness per Steves \etal~\cite{stevesCategorizingHumanPhishing2020}.
We performed measurements using the vendor's simulation platform, which reported the standard metrics directly and timing information that we used to derive the novel metrics.

\myparagraph{Standard Metrics}
We employed three standard behavioral metrics to assess training effectiveness~\cite{carellaImpactSecurityAwareness2017, caputoGoingSpearPhishing2014}:

\begin{itemize}[leftmargin=1.5em, itemsep=0pt, topsep=0pt]
  \item \textit{Open Rate} is the percent of recipients who \textit{open} the simulated phishing email, measured via embedded tracking pixels.\footnote{This is the standard definition, though we acknowledge it can be influenced by email client settings and security configurations~\cite{antoniouSystematicMethodExecute2015}.}
  \item \textit{Click-Through Rate (CTR)} is the percent of recipients who \textit{click} on malicious links within the phishing emails. 
  CTR is our primary indicator of phishing susceptibility~\cite{sumnerExaminingFactorsImpacting2022}.
  \item \textit{Reporting Rate} is the percent of recipients who \textit{report} the suspicious email to the organization's security team using the integrated ``Report Phish'' button, representing proactive security behavior and successful threat recognition~\cite{doingAnalysisPhishingReporting2024}.
\end{itemize}

\myparagraph{Novel Temporal Metrics}
Abroshan \etal suggested examining temporal patterns in user responses~\cite{abroshanPhishingHappensTechnology2021}, and Steves \etal asked: ``\textit{Is time to first report sooner than time to first click?}''~\cite{stevesCategorizingHumanPhishing2020}.
Rapid reporting of phish attempts enables the feedback loop built into modern email defenses (\cref{fig:PhishDefensesArchitecture}).
We introduce the following metrics to see organizational-level temporal security dynamics:


\begin{itemize}[leftmargin=1.5em, itemsep=0pt, topsep=0pt]
\item \textit{Individual Reporting Timeliness} measures how quickly individual users report suspicious emails:
$$\text{Timeliness}_i = t_{\text{report},i} - t_{\text{deployment}}$$
where $t_{\text{report},i}$ is the timestamp when user $i$ reported the email and $t_{\text{deployment}}$ is when the phishing campaign was launched.

\item \textit{Campaign Inoculation Status} determines if an organization can mitigate a phishing campaign before exploitation occurs:
$$\text{Inoculation} = \begin{cases} 
1 & \text{if } t_{\text{first report}} < t_{\text{first click}} \\
0 & \text{otherwise}
\end{cases}$$
where $t_{\text{first report}}$ and $t_{\text{first click}}$ are the timestamps of the first report and first click for a given campaign, respectively.

\item \textit{Organizational Inoculation Index (OII)} quantifies the temporal advantage (or disadvantage) of reporting over clicking:
$$\text{OII} = t_{\text{first click}} - t_{\text{first report}}$$
Positive values indicate ``inoculation'' where reports preceded clicks, enabling organizational threat mitigation before exploitation.
Negative values instead 
suggest vulnerability windows.
\end{itemize}

This novel organizational-level metric directly addresses Steves \etal's \cite{stevesCategorizingHumanPhishing2020} research question while providing actionable insights into collective security behaviors that complement individual-focused assessments.
The Organizational Inoculation Index lets organizations evaluate whether their workforce's collective reporting behavior provides early warning to enable proactive threat response.

\subsection{Analysis}
\label{sec:Methods-Analysis}

\myparagraph{Data Validation and Quality Assurance}
Prior to analysis, we conducted data validation to ensure analytical integrity.
This included verifying random assignment effectiveness across groups, assessing technical delivery success rates, and examining potential outliers or data anomalies~\cite{barrientosScalingPhishAdvancing2021}.

\myparagraph{Statistical Approach}
We employed two complementary analytical approaches to examine training effectiveness.
Our primary analysis used two-way analysis of variance (ANOVA) to test main effects and interactions between training modality (Treatments A, B, and Control) and phishing difficulty level (Easy, Medium, Hard) on our outcome measures.
Note that although ANOVA with binary outcomes violates standard normality assumptions, this approach aligns with established practices in phishing research~\cite{carellaImpactSecurityAwareness2017, caputoGoingSpearPhishing2014} and enables direct comparison with prior studies.

To address the limitations of ANOVA with binary data, we conducted confirmatory analyses using logistic regression models appropriate for binary outcomes (clicked/not, reported/not)~\cite{seabold2010statsmodels}. The logistic regression models employed maximum likelihood estimation and provided odds ratios for effect size interpretation.
Both analytical approaches yielded consistent conclusions regarding training effectiveness, with ANOVA providing F-statistics for comparison with prior literature and logistic regression offering more statistically appropriate modeling of our binary outcomes.
We report both approaches, while acknowledging the methodological trade-offs inherent in each approach.

\subsection{Threats to Validity}
\label{sec:Methods-Threats}

We encourage the reader to consider several limitations when interpreting the results of our study~\cite{verdecchia2023threats}.
We distinguish these limitations into three kinds of threats to validity.

\myparagraph{Construct Validity}
Construct validity is concerned with whether we operationalized the desired constructs of our study appropriately \cite{shadishCookCampbell2002,wohlinEtAl2012}. In this study, we worked with two primary constructs: (1) the phishing training we deployed, which depends on our cybersecurity training vendor; and (2) our assessment of the difficulty of phishing lures.
\textit{First}, though vendors may vary in training design, our cybersecurity vendor's training is regulation-compliant,  
and our results align with prior work that used different trainings.
\textit{Second}, to control for subjectivity in rating phishing lures, we used the NIST Phish Scale with multiple 
analysts to judge difficulty \cite{sjobergBergersen2023}.

\myparagraph{Internal Validity}
Internal validity threats affect the reliability of conclusions we have about the relationship between the cause and the effect we observe \cite{shadishCookCampbell2002,wohlinEtAl2012}. For our study, we used stratified random sampling to mitigate potential biases related to job role and other organizational issues.
Due to the large sample sizes for Treatments A and B, our study has significant statistical power.
Organizational constraints required us to limit the sample size for our control group.
This group was still large enough to be a baseline.

Like all real-world phishing measurement studies, our work faces several measurement concerns.
\textit{False positive clicks} may occur when automated security systems (security gateways, antivirus, etc.), ``shadow IT'' applications, or personally installed tools / browser extensions automatically scan or prefetch URLs from emails, generating click events without actual user interaction.
\textit{False negative reporting} may also occur, in which users report phishing emails through alternative channels, such as contacting the IT help-desk directly, rather than the simulation platform's designated ``Report Phish'' button.
These behaviors would, respectively, incorrectly increase and decrease our measurements of subject interactions.
We cannot fully eliminate such threats, but our large sample size mitigates their impact on our overall conclusions.

Although such measurement noise is inherent in operational studies, the large sample size and balanced distribution across conditions mitigate its systematic influence on our results.
Every effort was made to standardize user behavior, including explicit instruction during both training modalities on how to identify and report suspicious messages through the organization’s approved “Report Phish” mechanism.
Future work might explicitly quantify these instrumentation effects by correlating simulation telemetry with downstream SOC logs or ticketing data to better understand their prevalence and direction of bias.

\myparagraph{External Validity}
External threats impact the generalizability of our research~\cite{runesonHost2009,yin2018}.
Our study was situated in the USA at a fintech company.
The context of the fintech industry may limit generalizability to other sectors, as financial services employees may have a higher baseline level of security awareness. 
However, given the minimal effect sizes measured in our study, this threat is not substantial. 
Our focus on immediate post-training effects also limits our understanding of long-term training impact.
Future work might assess the effect of training recency and volume. 

\subsection{Ethics}
We acknowledge the importance of informed consent as part of an empirical study.
Employees are informed at onboarding that periodic phishing simulations are a mandatory condition of employment, consistent with industry practice in financial services organizations~\cite{doingAnalysisPhishingReporting2024}. 
Our study then analyzed data collected as part of routine, required security awareness training, using the existing corporate platform and controls. 

Our protocol was reviewed by both the company and the partner academic institution’s IRB.
Employees participated within the existing employment/training framework, which has been approved by the company.
In consultation with management, we excluded high-distress scenarios (\eg family emergency) to reduce psychological risk but preserve ecological validity.
Because the research used de-identified operational data generated by routine commercial operations, our institutional IRB concluded that from the academic partner's view our study did not constitute human subjects research. 

\section{Results}

\begin{table}
\centering
\caption{
Summary of Training Effectiveness Results.
Average rates show that trained groups performed similarly to control group, with modest differences across conditions.
}
\label{tab:ResultsSummary}
\footnotesize
\begin{tabular}{p{3.5cm}p{4cm}}
\toprule
\textbf{Aspect} & \textbf{Results} \\
\midrule
Effect Size & Consistent with large-scale findings~\cite{hoUnderstandingEfficacyPhishing2025} \\
\quad Training Effect & $\eta^2 = 0.000$ (negligible) \\
\quad Difficulty Effect & $\eta^2 = 0.007$ (meaningful) \\
\quad Training $\times$ Difficulty & $\eta^2 = 0.001$ (negligible) \\
\midrule
\multicolumn{2}{c}{\textbf{Click Rates}} \\
\midrule
Overall Click Rate & 10.4\% across all participants \\
\quad Control Group Average & 9.8\% \\
\quad All Trained Average & 10.5\% \\
\quad \quad Training Only & 10.6\% \\
\quad \quad Training + Exercise & 10.4\% \\
\midrule
\multicolumn{2}{c}{\textbf{Report Rates}} \\
\midrule
Overall Report Rate & 9.6\% across all participants \\
\quad Control Group Average & 8.9\% \\
\quad All Trained Average & 9.7\% \\
\quad \quad Training Only & 9.5\% \\
\quad \quad Training + Exercise & 9.9\% \\
\midrule
\multicolumn{2}{c}{\textbf{Statistical Effects}} \\
\midrule
Training Effect on Click Rates & p = 0.450 (not significant) \\
Training Effect on Report Rates & p = 0.417 (not significant) \\
NIST Phish Scale Effect & p $<$ 0.001 (highly significant) \\
\quad \textit{Easy} Difficulty & 7.0\% click rate \\
\quad \textit{Medium} Difficulty & 8.7\% click rate \\
\quad \textit{Hard} Difficulty & 15.0\% click rate \\
\midrule
Interaction Effects & Revealed nuanced patterns with difficulty levels \\
\bottomrule
\end{tabular}
\end{table}

Our analysis examined the effectiveness of phishing training across 12,511 participants randomly assigned to three experimental conditions.
\cref{tab:ResultsSummary} gives a summary of our key results.
\cref{fig:EffectSizeComparison} complements this with a depiction of effect sizes for the controlled factors.

    \label{fig:EffectSizeComparison}

\subsection{H1: Phish Scale Effect}

\begin{figure}
    \centering
    \includegraphics[width=1\linewidth]{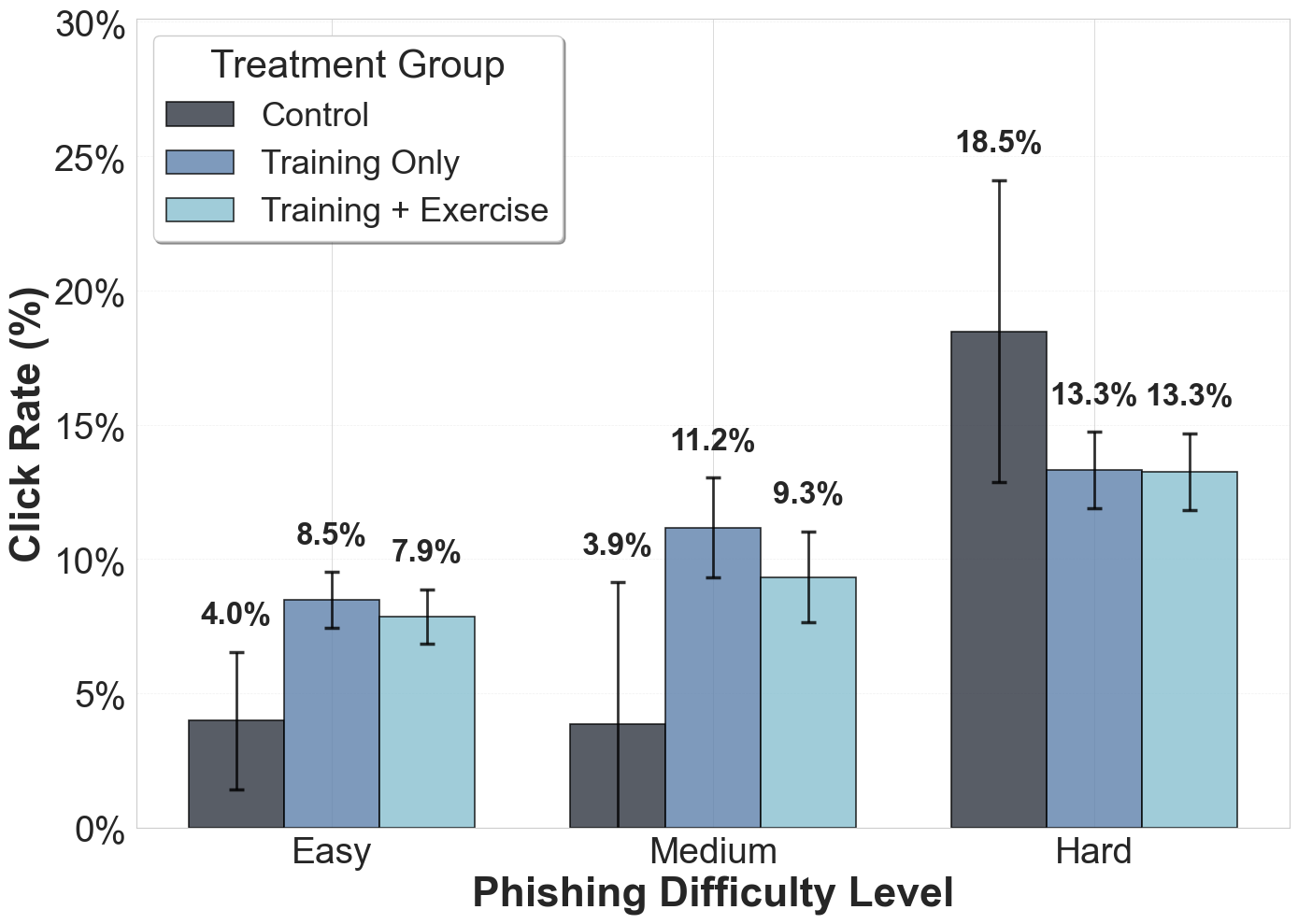}
    \Description{A bar chart comparing the effect sizes of phishing susceptibility factors.
    Each bar shows the variance explained by phishing difficulty, measured using the NIST Phish Scale,
    across the different training interventions. Difficulty accounts for most of the variance,
    while training modality and interaction effects are negligible.}    
    \caption{
Comparison of effect sizes of phishing susceptibility. Bars show the variance explained by phishing difficulty (per the NIST Phish Scale) for each training intervention.
}
    \label{fig:ResultsSummary}
\end{figure}

\textit{H1 was supported}.
\cref{fig:ResultsSummary} illustrates the comparison of effect sizes across different factors in phishing susceptibility.
The NIST Phish Scale successfully predicted user behavior. Phishing lure difficulty had a highly significant effect on click-through rates (F(2, 12086)=41.415, p\ $<$\ 0.001, $\eta^2$=0.007), validating the scale's utility for measuring phishing complexity. Higher difficulty emails consistently produced higher click rates across all experimental conditions; \textbf{Easy difficulty}: 7.0\% average click rate, \textbf{Medium difficulty}: 8.7\% average click rate, \textbf{Hard difficulty}: 15.0\% average click rate.


\subsection{H2: Overall Training Effects}

\textit{H2 was not supported}.
Training did not produce statistically significant improvements in either click reduction or reporting behavior. Main effect analysis showed no significant difference between trained and untrained groups for click rates (F(2, 12086) = 0.800, $p$=0.450) or report rates (F(2, 12086) = 0.874, $p$=0.417). This finding aligns with recent large-scale evidence suggesting minimal effectiveness of current training approaches~\cite{hoUnderstandingEfficacyPhishing2025}.

\subsection{H3: Interactive Training Effects}

\textit{H3 was not supported}.
Treatment B, interactive training (Training + Exercise) did not produce significantly higher reporting rates than traditional training alone.
While the interactive group showed numerically higher reporting rates across difficulty levels, these differences were not statistically significant. The coefficient estimate for Training + Exercise relative to Training Only was modest and non-significant across all difficulty levels.

This finding contrasts with some prior work suggesting benefits of interactive training components~\cite{carellaImpactSecurityAwareness2017}, but aligns with recent large-scale studies showing limited overall training effectiveness~\cite{hoUnderstandingEfficacyPhishing2025}.

\subsection{H4: Training-Difficulty Interactions}

\textit{H4 was not supported}. We found no statistically significant interaction between training modality and lure difficulty for click rates (F(4, 12086) = 3.135, $p$=0.014) or report rates (F(4, 12086) = 0.517, $p$=0.723). While there was a marginal interaction effect for click rates, this primarily reflected the pattern that training appeared to increase clicks on easier emails while potentially reducing clicks on harder emails.

Coefficient analysis revealed some interesting interaction patterns: \textbf{Training Only × Hard}: -9.67 percentage points ($p$=0.002); \textbf{Training + Exercise × Hard}: -9.09 percentage points ($p$=0.004)
These significant negative interactions suggest that training may provide some protection against the most difficult phishing emails, even though overall training effects were not significant. 
Higher alignment messages with fewer cues are the most deceptive~\cite{butaviciusBreachingHumanFirewall2015}.


\subsection{H5: Reporting Timeliness}

\textit{H5 was partially supported, though with important caveats about sample size limitations}.
Individual reporting behavior showed a median time-to-report of 21 minutes (0.35 hours), with 90\% of reports occurring within 18.8 hours of phishing deployment.
\JD{For individual reporting we should be able to measure the effect on the individual level, we have a large sample size for this. It is strange to report aggregated median and distribution instead.}
\DR{Checked the maths  - fixed the N of tests}
We analyzed 17 templates that received both clicks and reports. Overall, 52.9\% achieved inoculation (reports preceded clicks), with median Organizational Inoculation Index (OII) of +0.4 hours. By treatment group, Control achieved inoculation on 66.7\% of templates (median OII: +0.3 hours), Training Only on 75\% (median OII: +1.37 hours), and Training + Exercise on only 28.6\% (median OII: -0.2 hours). These differences were not statistically significant due to limited sample size. For successfully inoculated templates, the median OII was approximately 30 minutes, suggesting automated response mechanisms would be necessary to mitigate threats before exploitation.

These findings provide the first empirical measurement of organizational temporal security dynamics, responding to Steves \etal's research question about whether ``time to first report is sooner than time to first click''~\cite{stevesCategorizingHumanPhishing2020}.
The results suggest that organizational-level protective behaviors operate independently of individual training effectiveness.
Studying the opportunities and limitations of this feedback loop is a promising direction for further study. 




\subsection{Effect Sizes and Practical Significance}





While some coefficients reached statistical significance, effect sizes were generally small ($\eta^2$ $<$ 0.01), raising questions about practical significance~\cite{sumnerExaminingFactorsImpacting2022}.
The only meaningful effect was the difficulty gradient ($\eta^2$ = 0.007 for click rates), confirming the NIST Phish Scale's predictive validity. Even statistically significant interaction effects represented changes of less than 10 percentage points.

These findings underscore the need for realistic expectations about training effectiveness and suggest organizations should complement training with other defense mechanisms rather than relying on awareness programs as primary protection~\cite{franzSoK}.


\section{Discussion}


\subsection{Implications for Practice}


After deploying regulation-compliant training programs to over 12,000 employees, we found no statistically significant main effects of training on either click rates ($p$=0.450) or report rates ($p$=0.417).
The small effect sizes we observed ($\eta^2$ < 0.01 for all main effects) indicate that even statistically significant improvements translate to minimal operational impact. 
This result aligns with recent large-scale evidence from Ho \etal~\cite{hoUnderstandingEfficacyPhishing2025} and challenges the optimistic assessments promoted by cybersecurity training vendors.

The predictive power of the NIST Phish Scale (F(2, 12086)=41.415, p\ <\ 0.001) demonstrates that practitioners can significantly influence training results by adjusting the difficulty of phishing lures used in simulations. Our data shows click rates vary dramatically based on lure difficulty: from 7.0\% for easy emails to 15.0\% for hard emails. Organizations setting specific performance targets (such as click-through rates around 2\%) may achieve these by using low-difficulty lures that fail to represent sophisticated attacks organizations actually face~\cite{alkhalilPhishingAttacksRecent2021}. This creates a misalignment between perceived and actual security posture.

These findings suggest several actionable opportunities for both training vendors and organizational security teams.
Vendors can enhance their value by incorporating standardized measures of lure difficulty and adaptive testing frameworks aligned with models such as the NIST Phish Scale.
This would produce interpretable results that organizations could use to calibrate difficulty levels to track progress over time, personalize training to employee risk profiles, and benchmark outcomes across departments or industries.
However, the relatively low impact of phishing training suggests that awareness programs must be part of a layered defense strategy, complemented by technical controls and incident response preparedness~\cite{franzSoK}, to gain risk reduction rather than nominal compliance~\cite{haneyComplianceImpactTracing2024}.
In this way, security awareness initiatives can evolve from static, one-size-fits-all modules into data-driven, context-aware components of an organization’s continuous improvement process.

\subsection{Future Work}

\myparagraph{Refinement of Organizational Inoculation Index}
Our introduction of the Organizational Inoculation Index offers a starting point for measuring collective security behaviors across a workforce.
We suggest that this metric might be a useful addition to large-scale phishing simulation tools, as it provides a valuable assessment of how quickly and effectively an organization responds to bulk phishing threats at scale, although its utility against spear phishing (especially in the age of LLMs) is unclear. 

\myparagraph{AI-Generated Phishing and the Limits of the NIST Phish Scale}
The emergence of generative AI enables highly convincing phishing attacks at scale that lack traditional flaws used to assess difficulty, such as unsecured websites~\cite{kim2021sooel} or poor grammar~\cite{royFromChatbotsPhishbots2024}. This calls into question the continued relevance of the NIST Phish Scale. Future research should investigate whether training methods can prepare users to recognize AI-crafted phishing attempts.

\section{Conclusion}

This work provides real-world evidence on the efficacy of phishing awareness training and offers the largest validation to date of the NIST Phish Scale through its application under multiple experimental conditions. Difficulty strongly predicted user behavior (click rates from \(7.0\%\) for easy lures to \(15.0\%\) for hard; \(F(2,12086)=41.415,\, p<0.001\)), while neither interactive nor lecture-based training produced statistically significant changes in click rates ($p$=0.450) or reporting behavior ($p$=0.417). Effect sizes for all training effects were minimal. These results confirm current research and run counter to industry assumptions and marketing claims regarding the efficacy of training tools.
Our evidence indicates that training provides only modest protection against evolving phishing threats and should be regarded as just one component of a broader security awareness program.
Our results reproduce those of prior work while evaluating new hypotheses and a new context.


\bibliographystyle{plain}
\bibliography{bib/phish}

\appendix

\section*{Appendices}

The extended material presented here consists of:

\begin{itemize}
    \item \Cref{sec:appendix:EmailControlsDetail}: The studied organization's email flow
    \item \Cref{sec:appendix-NISTPhishScale}: Definition of the NIST Phish Scale
    \item \Cref{sec:appendix-Lures}: Details of the lures used
\end{itemize}

\section{Email Controls in the Studied Organization}
\label{sec:appendix:EmailControlsDetail}

\begin{figure*}[!t]
    \centering
    \includegraphics[width=0.9\linewidth]{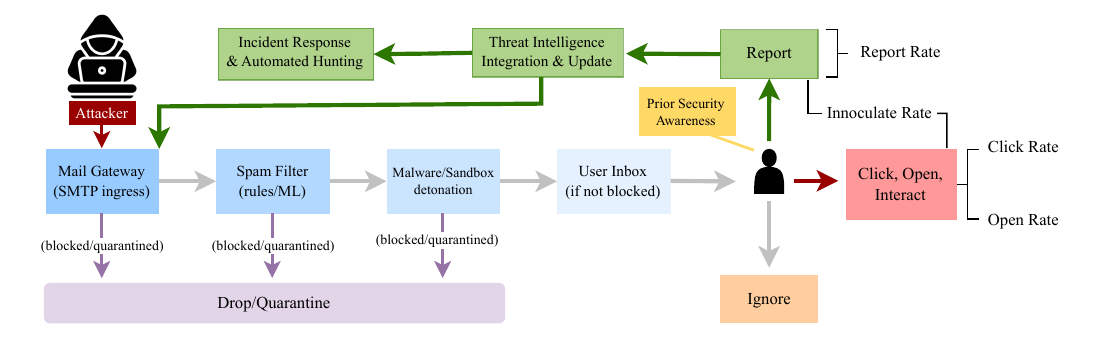}
    \Description{A systems diagram of organizational phishing defenses.
    The flow begins with inbound email entering through SMTP gateways and spam filters,
    then passes through malware and sandbox analysis before reaching user inboxes if not blocked.
    Users can either click or report the message.
    Reports feed into incident response and threat intelligence systems,
    which update filters and quarantines to protect others in the organization.}
    \caption{
    Phishing defense architecture, mapping the email pathway from authentication through post-delivery response.
    Technical controls create multiple interception points prior to human interaction, and human reports provide a feedback opportunity for organizational-level threat mitigation.
    }
    \label{fig:PhishDefensesArchitecture}
\end{figure*}

\cref{fig:PhishDefensesArchitecture} depicts the studied organization's email processing flow, with defenses noted and our metrics placed in the broader context.
 
 The following is a detailed sequence of the controls in the studied organization's email defense setup.

 \begin{description}[align=left, leftmargin=0.5em, labelsep=0.5em]
   \item[Malicious Email Sent] An attacker crafts a convincing message and sends it from a look-alike or compromised account, often using links or attachments that lead to theft or malware \cite{nist800177r1-2020}.
   \item[Email Authentication: SPF] The domain owner lists which mail servers are allowed to send on its behalf; receivers check the sender's IP against that list to spot forgeries \cite{rfc7208spf2014}.
   \item[Email Authentication: DKIM] The sender adds a digital signature to the message; receivers use a public key in DNS to confirm the message really came from that domain and wasn't altered \cite{rfc6376dkim2011}.
   \item[Email Authentication: DMARC] Builds on SPF/DKIM to tell receivers what to do with failures (monitor, quarantine, or reject) and sends reports back to the domain owner \cite{rfc7489dmarc2015}.
   \item[Email Authentication: BIMI] A visual cue: providers that support it can display a verified brand logo when strong DMARC is in place; it helps users spot impostors \cite{bimi-group,bimi-ietf-guidance-2025}.
   \item[Email Authentication: ARC] Preserves original authentication results across forwarding/mailing lists so legitimate forwarded mail isn't penalized \cite{rfc8617arc2019}.
   \item[Transport Encryption: STARTTLS] Upgrades the SMTP connection to TLS so email isn't readable in transit between servers \cite{nist800177r1-2020}.
   \item[Transport Enforcement: MTA-STS] Lets a domain publish a policy that its email must use TLS with valid certificates—blocking downgrade/MITM attempts \cite{rfc8461mtasts2018}.
   \item[Transport Enforcement: DANE] Uses DNSSEC to bind TLS certificates to the domain for SMTP, strengthening spoofing/interception resistance \cite{rfc7672dane2015,rfc6698dane2012}.
   \item[Secure Email Gateway (SEG)] The organization's front door for mail: filters spam/phishing, checks sender/domain reputation, analyzes headers/content, and often rewrites links for time-of-click checks \cite{nist800177r1-2020,msft-safe-attachments-2025}.
   \item[Threat Intelligence Integration \& Update] Feeds of known-bad IPs, domains, URLs, and file hashes; standards like STIX/TAXII and programs like CISA AIS for sharing/automation \cite{oasis-stix21,oasis-taxii21,cisa-ais}.
   \item[AI/ML Threat Detection] Behavioral and language models look for impersonation and anomalies beyond fixed indicators—useful for brand-new or targeted lures \cite{google-tensorflow-2019,google-ai-spam-2024}.
   \item[Attachment Sandboxing] Opens attachments in an isolated virtual environment (“detonation”) to observe malicious behavior and block delivery \cite{msft-safe-attachments-2025}.
   \item[Content Disarm \& Reconstruction (CDR)] Strips active content (macros, scripts) from files and rebuilds a safe version so users can still open the content \cite{nist800177r1-2020}.
   \item[Data Loss Prevention (Outbound Scanning)] Watches outgoing mail for sensitive data (\eg customer lists, PHI); can also flag compromised accounts trying to exfiltrate information \cite{msft-purview-dlp-learn-2025}.
   \item[Delivered to User Inbox] Messages that pass earlier controls arrive for the user—this is where human judgment and endpoint protections matter \cite{nist800177r1-2020}.
   \item[Email Client Security] Clients/services block auto-loading images, scan/contain attachments, and add safety banners to reduce risk on open/click \cite{gmail-safety-center,outlook-block-images}.
   \item[Security Awareness Training \& Simulation] Short, regular training and realistic simulations teach users what to look for and build a habit of reporting \cite{nist80050r1-2024}.
   \item[User Reports Phish] One-click ``Report Phish'' sends the message to security; reported samples help purge look-alikes and improve filters \cite{gmail-report-phish,msft-report-message}.
   \item[SIEM/SOAR Integration] Centralizes mail/endpoint alerts and runs automated playbooks to block senders, remove messages, and open tickets \cite{cis-control8-2025,nist80061r3-2025}.
   \item[Incident Response \& Automated Hunting] Analysts confirm threats, retro-hunt for variants, remove messages, and contain affected accounts/devices \cite{nist80061r3-2025,msft-campaigns-2025}.
   \item[Zero-Hour Auto Purge (ZAP)] If a campaign is discovered post-delivery, the system can retroactively remove those emails from mailboxes \cite{msft-zap}.
   \item[Email Quarantine Management] Suspicious but unconfirmed messages are held for review so security teams (or users, with guardrails) can release or delete \cite{msft-quarantine,gmail-quarantine}.
   \item[Mobile Device Management (MDM)] Enforces secure mail apps, screen locks, and remote wipe on phones/tablets that access corporate email \cite{nist800124r2-2020}.
   \item[Email Retention \& Backup Systems] Archiving/backups keep searchable records for investigations/compliance—and let you restore mailboxes after an incident \cite{nara-capstone,exo-archiving}.
 \end{description}

\section{Measuring Lure Complexity with the NIST Phish Scale}
\label{sec:appendix-NISTPhishScale}

As discussed in~\cref{subsec:empirical-studies}, we used the NIST Phish Scale as a standardized measure of phishing lure difficulty, \ie the relative difficulty of detecting whether or not a given email is part of a phishing attack.
The NIST Phish Scale is intended to help organizations benchmark lure complexity and contextualize simulation outcomes~\cite{canhamPhishDerbyShoring2022}.
It categorizes lure difficulty along two dimensions, shown in~\cref{tab:phish-scale-matrix}.
Barrientos \etal~\cite{barrientosScalingPhishAdvancing2021} conducted early validation work on the NIST Phish Scale, while Dawkins and Jacobs~\cite{dawkinsScalingNISTPhish2023} provided the official implementation guidelines.
Ours is the first study to validate the Phish Scale's premise at scale in a real-world context.


\begin{table}[h]
\centering
\small
\caption{
NIST Phish Scale for Detection Difficulty~\cite{stevesCategorizingHumanPhishing2020,dawkinsScalingNISTPhish2023}.
\textit{Phishing Cues} represent observable errors or inconsistencies in the email that might alert users to its fraudulent nature, \eg suspicious URLs or formatting irregularities~\cite{canhamNotAllVictims2024}.
\textit{Premise Alignment} measures the relevance of the email to the recipient's organizational context~\cite{antoniouSystematicMethodExecute2015}.
}
\label{tab:phish-scale-matrix}
\begin{tabular}{l|ccc}
\toprule
\textbf{Cues {\textbackslash} Alignment} & \textbf{Weak} & \textbf{Medium} & \textbf{Strong} \\
\toprule
\textbf{Few} & Medium & Hard & Hard \\
\textbf{Some} & Easy-to-Medium & Medium & Hard \\
\textbf{Many} & Easy & Medium & Medium \\
\bottomrule
\end{tabular}
\end{table}

\begin{table*}
\centering
\caption{
NIST Phish Scale Assessment of Phishing Templates Used In This Study.
Premise Alignment scores range from 1 (Low) to 3 (High).
Lure Scores range from 1-3 based on phishing cue assessment.
Cue Count represents the total number of suspicious indicators identified in each template.
Based on lure and premise scores, we mapped the top group to ``Easy'', the middle group to ``Medium'', and the bottom group to ``Hard''.
}
\label{tab:phish-scale-assessment}
\footnotesize
\small
\begin{tabular}{p{3cm}ccccp{5.5cm}}
\toprule
\textbf{Template Name} & \textbf{Difficulty} & \textbf{Cue Count} & \textbf{Lure Score} & \textbf{Premise Score} & \textbf{Premise Alignment Explanation} \\
\midrule
Attachment - XLS & Easy & 14 & 2 & 2 & Moderate alignment: XLS financial data plausible, but lacks sender/context. \\
Fax via Voicemail Office & Easy & 13 & 2 & 1 & Low alignment: Fax delivery uncommon in modern workplace. \\
Email Not Delivered | Office & Easy & 13 & 2 & 3 & High alignment: Bounce messages routine and widely expected. \\
Email Not Delivered | Google & Easy & 16 & 3 & 3 & High alignment: Gmail bounce notices with technical detail believable. \\
From Marketing: Collab Request & Easy & 14 & 2 & 3 & High alignment: Internal collaboration messages from marketing common. \\
Accounting Team Amex Charges | Office & Easy & 16 & 3 & 3 & High alignment: Finance inquiries during month-end familiar. \\
HSA Use Balance! & Easy & 17 & 3 & 2 & Moderate alignment: HSA reminders expected, but lacks personalization. \\
Tax Workshop & Easy & 16 & 3 & 2 & Moderate alignment: HR tax webinars plausible but weakened by generic formatting. \\
\midrule
Device Non-compliant - Microsoft & Medium & 15 & 3 & 3 & High alignment: Matches expected MDM behavior and terminology. \\
Mobile Device Restricted | Outlook & Medium & 16 & 3 & 3 & High alignment: Device quarantine messages match common security flows. \\
Device Non-compliant - Google & Medium & 16 & 3 & 3 & High alignment: Technical restriction notices align with enterprise environments. \\
\midrule
Attachment - Word & Hard & 14 & 2 & 2 & Moderate alignment: Document sharing plausible, but placeholder fields reduce realism. \\
Vulnerabilities | Office & Hard & 18 & 3 & 1 & Low alignment: Scenario vague and not role-specific; unlikely context. \\
Loom - Recording Shared & Hard & 14 & 2 & 2 & Moderate alignment: Loom recordings common, but template fields weaken trust. \\
Mentor Program & Hard & 15 & 3 & 3 & High alignment: Internal mentorship invitations believable and well-framed. \\
HR - Performance Review | Outlook & Hard & 16 & 3 & 3 & High alignment: Outlook calendar invites for reviews routine. \\
Changes to Healthcare Form & Hard & 15 & 3 & 3 & High alignment: Healthcare form updates common in HR communications. \\
Vulnerabilities | Google & Hard & 16 & 3 & 1 & Low alignment: Scenario lacks specificity and role-relevant context. \\

\bottomrule
\end{tabular}
\end{table*}
\section{Lures}
\label{sec:appendix-Lures}

See~\cref{tab:phish-scale-assessment} for our analysis of the phishing lures used in this study.

Following the NIST Phish Scale framework, we created a 3×3 classification matrix combining cue visibility (few-low, some-medium, many-high) with premise alignment (weak-low, medium, strong-high). Templates were assigned composite difficulty scores by summing both dimensions, with scores ranging from 2 (easy: high cues, low alignment) to 6 (hard: low cues, high alignment). Based on lure and premise scores, we mapped the top group to ``Easy'', the middle group to ``Medium'', and the bottom group to ``Hard''.

The deployed templates were constrained by organizational considerations.
ACME Corp.'s management requested the exclusion of highly distressing content (\eg emergency family situations, personnel security scenarios) to minimize employee psychological impact.
This constraint reflects the practical limitations of conducting research within operational environments, where employee welfare considerations influence experimental designs~\cite{doingAnalysisPhishingReporting2024}.

\end{document}